\documentclass[prl,twocolumn,superscriptaddress,doublespace]{revtex4}
\pdfoutput=1
\usepackage{graphicx}
\usepackage{amsmath}
\usepackage[usenames]{color}
\usepackage{amssymb}
\newcommand{\be}{\begin{equation}}
\newcommand{\ee}{\end{equation}}
\newcommand{\bea}{\begin{eqnarray}}
\newcommand{\eea}{\end{eqnarray}}

\newcommand{\nno}{\nonumber}
\newcommand{\bse}{\begin{subequations}}
\newcommand{\ese}{\end{subequations}}

\begin{document}
\title{Minimal Higgs inflation}
\author{Debaprasad Maity \footnote{debu.imsc@gmail.com}}
\affiliation{Department of Physics, 
Indian Institute of Technology, Guwahati, India}

\begin{abstract}
In this paper we propose two simple minimal Higgs inflation scenarios through a simple modification of the Higgs potential,
as opposed to the usual non-minimal Higgs-gravity coupling prescription.
The modification is done in such a way that it creates a flat plateau for a huge range of field values 
at the inflationary energy scale $\mu \simeq (\lambda)^{1/4} \alpha$. Assuming the perturbative Higgs quartic coupling, $\lambda \simeq {\cal O}(1)$, for both the models inflation 
energy scale turned out to be $\mu \simeq (10^{14}, 10^{15})$ GeV, and prediction of all the cosmologically 
relevant quantities, $(n_s,r,dn_s^k)$, fit extremely well with observations made by PLANCK.
Considering observed central value of the scalar spectral index, $n_s= 0.968$, our two models predict
efolding number, $N = (52,47)$.  Within a wide range of viable parameter space, we found that the prediction of 
tensor to scalar ratio $r (\leq 10^{-5})$ is far below the current experimental sensitivity to be observed in the near future.
The prediction for the running of scalar spectral index, $dn_s^k$, approximately remains the same as was predicted by the usual 
chaotic and quartic inflation scenario.  We also the computed the background field dependent unitarity scale $\Lambda(h)$, which turned out to be
much larger than the aforementioned inflationary energy.

\end{abstract}

\maketitle

\newpage
Higgs inflation is an interesting model proposed long time back \cite{shaposnikov} in order to
creat a bridge between the two most successful standard models in physics.
In one hand we have the standard model of particle physics which has been
experimentally verified to an extremely great precession. On the other hand,
standard model of cosmology reached to a level where the robustness is just an undeniable fact.
However, both models have one striking similarity at their respective high energy scale of interest at which extensive research
are still going on. In particle physics, the Higgs mechanism is an integral part, which
is controlled by a scalar field called Higgs, which has been recently discovered \cite{higgs1,higgs2}.
However, properties of this field is not very well understood yet. 
Similarly, in the cosmology, the inflationary mechanism \cite{Guth,steinhardt,linde} is also believed to be an integral part,
which can be best explained by incorporating a scalar field called inflaton. However,
the way, we have understood the Higgs because of its experimental accessibility, is unlikely 
for the inflaton case. Because of the huge energy gap between the TeV scale Higgs physics
and usually nearly GUT scale inflaton physics, it is hard to belive the existence of any connection
between the two. Any endeavour towards making this identification would be
very challenging from the effective field theory point of view. The main challenge to realise the minimal Higgs inflation without
any further modification, is the lack of tuning parameter. All the parameter associated with the Higgs field
are fixed by the TeV scale physics such as dimensionless Higgs quartic coupling, $\lambda h^4$,
is constrained as $0.11 < \lambda <0.27$ \cite{olive}. On the other hand, it has been shown that one requires 
$\lambda \leq 10^{-9}$ to produce right magnitude of density fluctuation during inflation. 
Therefore, to circumvent this problem, the very first attempt towards 
this direction was made by \cite{shaposnikov}, and subsequently 
various other roots have been taken \cite{CervantesCota:1995tz,Germani:2010gm,Kamada:2010qe} to identify
Higgs as an inflaton field. The main ingredient of all those inflation scenarios
is a non-minimal Higgs-gravity coupling. However, immediately after the proposal, all those models
have been questioned considering unitarity issue \cite{Burgess,Lerner,Atkins:2010yg}.

In this letter we propose a new Higgs inflation scenario, where instead of introducing non-minimal coupling with the gravity
we rather modify the potential at the ultra-violate regime in such a way that the TeV scale contribution of those modifications
could be suppressed by a new scale $\alpha$. However, full quantum field theory analysis is required to verify this. 
Through naive dimensional analysis of operators, we consider the following two possible form of the Higgs quaric potentials,
\bea
V_{h} =\left\{\begin{array}{cc} \frac {\lambda}{4} \frac{({\bf H}^{\dagger} {\bf H} - v^2)^2}{1 + \frac{({\bf H}^{\dagger} {\bf H})^2}{\alpha^4}} \\ 
\frac {\lambda}{4} \frac{({\bf H}^{\dagger} {\bf H} - v^2)^2}{\left(1 + \frac{({\bf H}^{\dagger} {\bf H})}{\alpha^2}\right)^2} .
\end{array}\right.
\eea   
Where, ${\bf H}$ is the SU(2) Higgs doublet. From the experiment we know $\lambda \simeq 0.11-0.24$ and
Higgs vacuum expectation value $v = 246$ GeV \cite{olive}. For the each potential form, the value of new scale $\alpha$ 
will set the scale of inflation. Therefore, both the potentials will contribute 
an infinite series of $\alpha$ suppressed operators at the perturbative label at TeV scale.
From the low energy quantum field theory point of view, naive power counting analysis also suggests that
the unitarity may be violated at scale $\alpha$. However, it may not be a problem for our proposal,
since the inflationary dynamics includes infinite series of higher dimensional operators. 
In addition another important fact is that the background dependent cut unitarity scale $\Lambda(h)$
turned out to be much large than the above tree level unitarity scale i.e. $\Lambda(h) \gg \alpha$.
Detailed analysis of this issue should be done before we get to any conclusion. Studying the phenomenological aspects
of those operators could also be important at TeV scale. Nevertheless, the basic requirement of any Higgs inflation 
scenario is to reproduce all cosmological quantities considering the parameters $(\lambda, v)$ at their TeV scale value.
Hence, our goal is to pin down the value of the new scale $\alpha (\gg v)$ in consistent with
the cosmological observation made by PLANCK \cite{PLANCK}. Since, we are interested in inflation, we can take the real component
of the Higgs field $h$, so that the inflationary potential will turn out to be
\bea
V_{h} =\left\{\begin{array}{cc} \frac {\lambda}{4} \frac{h^4}{1 + \frac{h^4}{\alpha^4}} \\
\frac {\lambda}{4} \frac{h^4}{\left(1 + \frac{h^2}{\alpha^2}\right)^2} .
\end{array}\right.
\eea   
Therefore, we start with the following Einstein-Hilbert action with the minimally coupled
Higgs field,   
 action, 
\be {\cal S}_H=\int
d^4x\sqrt{-g}[\frac 1 2 M_p^2 R-\frac{1}{2}
\partial_\mu h \partial^\mu h- V_{h}]~,
\ee
 where
$M_p = 1/\sqrt{8\pi G} = 2.45 \times 10^{18}$ GeV, is the reduced Planck mass. $R$ is the Ricci scalar. 
Considering the usual FRW metric, 
\be\label{metricfrw} ds^2=-dt^2 + a^2(t)(dx^2+dy^2+dz^2)~,\ee 
the equations of motion for the metric and the
Higgs field $h$ are: 
\be 3 M_p^2 H^2=\left(\frac{1}{2}\dot h^2+ V_{h}\right)~,~~~\ddot{h}+3H\dot{h}+V'_{h}=0~.
\ee

In order to achieve sufficient number of efolding, we identify various "slow-roll" parameters
as follows,
\be
\epsilon = \frac {M_p^2}{2} \left(\frac{V'_{h}}{V_{h}}\right)^2~;~
\eta = M_p^2 \frac{V''_{h}}{V_{h}} ~;~\xi = M_p^4 \frac{V'_{h}V'''_{h}}{V_{h}^2} 
\label{slow}
\ee 
During the inflation, all those parameters has to be much smaller than unity.
Therefore, violation of smallness of any one of those slow roll parameters will end the 
inflationary dynamics. Usually we take $\epsilon \simeq 1$ to identify the end of inflation. 
To set the beginning of inflation, we define another important cosmological quantity called 
efolding number, which measures the amount of inflation required to explain our observed universe. 
The expressions for the efolding numbers are,
\bea
N =\left\{\begin{array}{cc} \frac{\alpha^2}{4M_p^2} \left(\frac{(\tilde{h}^6 - 
\tilde{h}_{end}^6)}{6} + \frac{(\tilde{h}^2 - \tilde{h}_{end}^2)}{2}\right)\simeq \frac {\alpha^2}{24 M_p^2} {\tilde h}^6  \\
 \frac{\alpha^2}{4M_p^2} \left(\frac{(\tilde{h}^4 - \tilde{h}_{end}^4)}{4} + 
\frac{(\tilde{h}^2 - \tilde{h}_{end}^2)}{2}\right) \simeq \frac {\alpha^2}{16 M_p^2} {\tilde h}^4 .\\
\end{array}\right.
\label{efold}
\eea   
Where, we have defined dimensionless field $\tilde{h} = h/\alpha$, and the suffix "end" corresponds
to the value of Higgs field at the end of inflation, specifically at the point where the slow 
roll parameter $\epsilon \simeq 1$. Therefore, depending upon the requirement
of the number of efolding, we get the initial value of the inflaton.
 In this 
letter we will consider $N=(50,60)$, and their predictions for the inflationary 
observables $(n_s,r)$. The scalar spectral index $n_s$ and the tensor to
scalar ratio $r$ are related to the slow roll parameters we have defined earlier eq.(\ref{slow}) as
follows:
\bea
&&1-n_s = 6 \epsilon - 2 \eta \simeq \left\{\begin{array}{cc} \frac {5} {3 N} \\ \nno
  \frac {3} {2 N}
\end{array}\right. \\
&&r = 16 \epsilon \simeq \left\{\begin{array}{cc} \frac {4}{3^{\frac 5 3}} \left(\frac{\alpha}{M_p}\right)^{\frac 4 3} \frac {1}{N^{\frac 5 3}}\\ 
  \frac{2\alpha}{M_p} \frac {1}{N^{\frac 3 2}}
\end{array}\right.
\eea

\begin{figure}
\begin{center}
  \includegraphics[width=008.70cm,height=03.142cm]{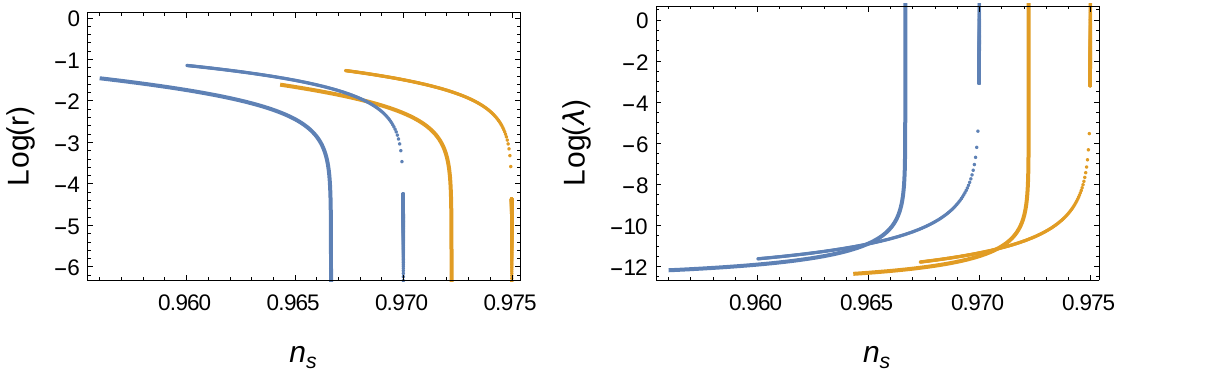}
   \caption{\scriptsize In the left we plotted $(n_s~ vs~ \log{\lambda})$, and in the right we plotted
$(n_s ~vs~ \log{r})$. We have considered $N=(50,60)$. For all the plots, blue curves are for $N=50$,
and orange curves are for $N=60$. The solid curves are for the first model and the dotted cures are 
for the second model. Every point on a particular curve corresponds to a particular value of inflationary energy scale $\alpha$. We considered 
$\alpha$ within the range $(0.0001, 100)$ in Planck unit. As one decreases the value of $\alpha$ each models has stabilized values 
of $n_s$. However, the value of $(r, \lambda)$ become
sensitive at lower value of $\alpha$.}  
\label{N5060}
\end{center}
\end{figure}

For the above final expressions, we have used the condition $\alpha < M_p $, and also ignored the
contribution from the $h_{end}$ in eq.(\ref{efold}). In the expression for spectral index, one sees that the
leading order contribution is coming from slow roll parameter $\eta$, whose leading behaviour does not depend on $\alpha$.
We also numerically checked our claim. Important to notice that the predictions of $r$ for both form of the potentials
turned out to be very small ($\ll 0.11$) in consistent with the PLANCK result. We also see this fact in the left panel of Fig.(\ref{N5060}), specifically the vertical part of the curves. Associated with the efolding number $(N)$, and the tensor to scalar ratio $(r)$, 
a quantity of theoretical interest called Lyth bound \cite{lyth}, $\Delta h$, is defined. This quantity tells us the maximum possible
value that the inflaton can travel during inflation. From the effective field theory point view, the Lyth bound can
question the effective validity of a particular model under consideration at an energy scale of interest. The bound
on this field excursion $\Delta h$ can be expressed as,
\bea
{\Delta h} \gtrsim ~N M_p \sqrt{\frac{r}{8}} = \left\{\begin{array}{cc}  \frac {M_p}{3^{\frac 5 6}} \left(\frac{\alpha}{M_p}\right)^{\frac 2 3} {N^{\frac 1 6}}\\ 
  \frac{M_p}{\sqrt{2}} \left(\frac{\alpha}{M_p}\right)^{\frac 1 2} N^{\frac 1 4} .
\end{array}\right.
\label{deltaphi}
\eea
As one can immediately see that $\Delta h$, contains the parameter $\alpha$ with a positive power. Therefore, we 
will see, our minimal model naturally gives sub-Planckian field excursion if we demand the Higgs quartic coupling to be of order
unity.  
At this point we would like to point out that, our minimal Higgs model does not belong to
the class of recently proposed supergravity inspired inflationary $\alpha$ attractor \cite{attractor} models,
as those models predict $1-n_s \simeq 2/N, r \simeq 12\alpha/N^2$, for $\alpha \leq {\cal{O}}(10)$ and large N. Interestingly,
the usual Higgs inflation model \cite{shaposnikov} is one of the members in that class of models. 
Therefore, it would interesting to find out supergravity origin of our models. 
Another interesting observable which also can constrain
various inflationary models is the running of spectral index, which can be expressed as
\bea
dn_s^k = \frac{dn_s}{d \ln k} =- 24 \epsilon^2 + 16 \epsilon \eta - 2 \xi \simeq \left\{\begin{array}{cc} \frac {-5} {6 N^2} \\ 
  \frac {-3} {4 N^2} .
\end{array}\right.
\eea
In the above expression, we only keep the leading order in $N$. Important to see that, it does not depend 
upon the value of $\alpha$. Therefore, we will have definite but small prediction for the running of scalar spectral index.
To the lading order in $N$, the behaviour of spectral running is same as the usual chaotic inflation.   
We also have numerically checked the validity of all the approximations made in the above expressions. 
So far all the quantities we discussed are independent of $\lambda$. The constraint on $\lambda$ can be obtained from the 
\cite{Stewart:1993bc}, the primordial power spectrum of the curvature perturbation: 
\be\label{spectrum} {\cal
P}_\zeta = \left\{\begin{array}{cc} \frac {\lambda \alpha^6 \tilde{h}^6 (1+ \tilde{h}^4)}{12\times 64 \pi^2 M_p^6} 
\simeq \frac {\lambda}{4 \pi^2} \left(\frac{\alpha}{M_p}\right)^{\frac 8 3} N^{\frac 5 3} \\
  \frac {\lambda \alpha^6 \tilde{h}^6}{12\times 64 \pi^2 M_p^6} \simeq \frac {\lambda}{12 \pi^2} \left(\frac{\alpha}{M_p}\right)^{3} N^{\frac 3 2}  \\  
\end{array}\right.,
\ee  
where, $\zeta$ is the variable for gauge invariant scalar perturbation,
Through out our analysis, we will be considering PLANCK central value of
$n_s = 0.9682 \pm 0.0062$ and the normalization for ${\cal P}_{\zeta} = 2.4 \times 10^{-9}$
at the pivot scale $k = 0.005 Mpc^{-1}$. For the minimal Higgs like 
quartic potential, field value takes $h \geq M_p$, in order to achieve sufficient 
number of efoling, and correct amplitude of density perturbation leads to an unnatural 
value of the Higgs quartic coupling i.e $\lambda \leq 10^{-9}$. This is believed to be an unnatural 
value for a dimensionless number, which means radiatively unstable from the quantum field
theory point of view. More importantly, it leads to a huge mismatch with the
standard model prediction, $0.11<\lambda \lesssim0.27$ \cite{olive}. 
As mentioned before, to circumvent this problem, the first model was proposed in \cite{shaposnikov} by introducing 
a non-minimal coupling of the Higgs with the Ricci scalar such as 
${\cal L}_{int} \sim \xi h^2 R$, where $\xi$ is the dimensionless coupling. The value
of $\xi (\geq 44700 \sqrt{\lambda})$ is very high to maintain the flatness of the potential during inflation.
However, the model has been questioned, as unitarity \cite{Burgess,Lerner} may violate much before the inflationary 
energy scale.
\begin{figure}
\begin{center}
  \includegraphics[width=006.40cm,height=04.52cm]{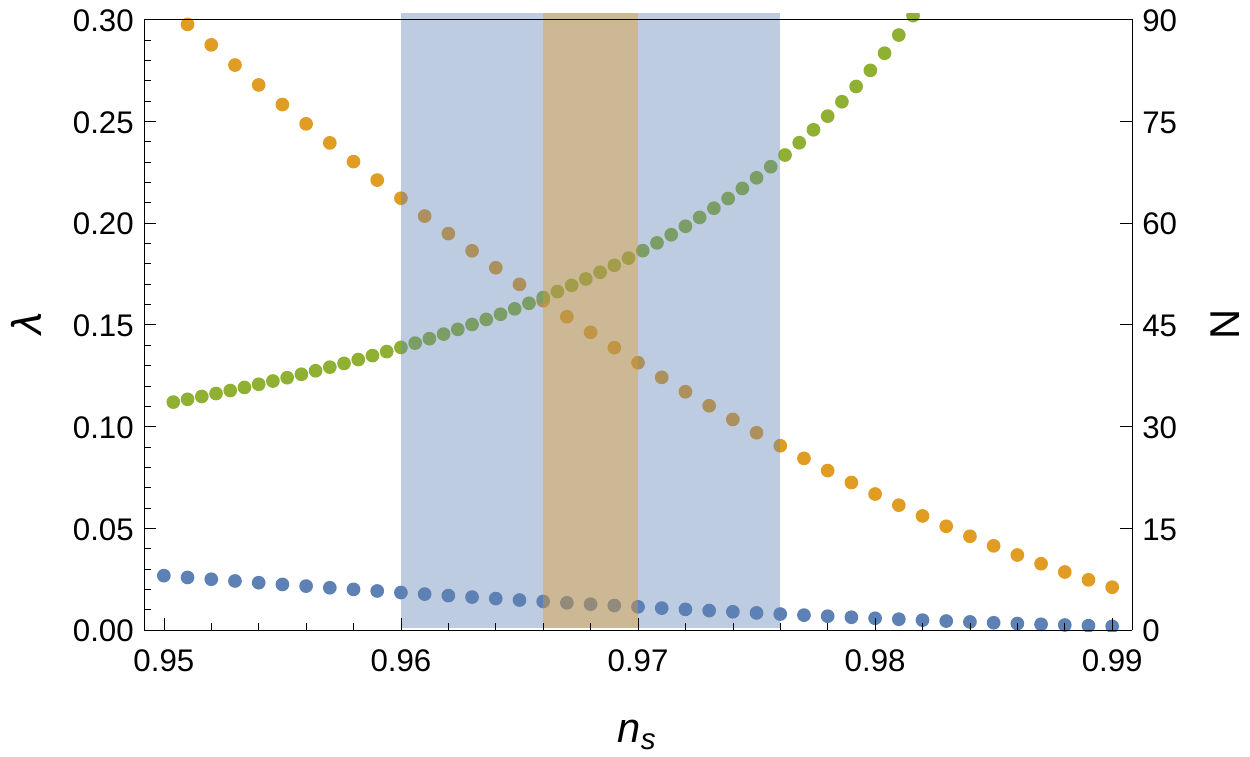}
   \caption{\scriptsize On the same figure, we have plotted $(n_s~vs~ \lambda)$ and $(n_s ~vs~ N)$, for two different values of $\alpha=0.001, 0.0004$ 
in Planck unit. a) ${\bf (n_s~vs~\lambda)}$ plot: The ({\bf blue},{\bf orange}) dotted curves are
for ${\alpha}/{M_p}=0.001, 0.0004$ respectively. b)  ${\bf (n_s ~vs~ N)}$ plot: The {\bf green} dotted curve is for both $\alpha$ values, as both turned out to be
superimposed. This plot is for the first model. The behaviour will be same for second model with the value of $\alpha$, which is one order of magnitude less compared to
value we considered for the first model. We also showed the bounds on $n_s$ from PLANCK.  $1 \sigma$ bounds on $n_s$, corresponds to 
the light blue shaded vertical region. $1 \sigma$ bounds of a further CMB experiment with sensitivity $\pm 10^{-3}$ \cite{limit1,limit2}, corresponds to
brown shaded region. For all those bounds, the central value $n_s =0.9682$ is considered.   
}
\label{lambda}
\end{center}
\end{figure}
However, subsequently in the reference \cite{shaposnikov2}, the author introduced the concept of background field
dependent cut off, so that the unitarity scale could be very high compared to the previous claim. 
In an another model, a more complicated, Einstein tensor $G^{\mu\nu}$ 
coupled with the kinetic term of the Higgs field, $G^{\mu\nu}\partial_\mu h\partial_\nu h$
\cite{CervantesCota:1995tz,Germani:2010gm} has been considered, but claimed to be plagued with the 
unitarity problem \cite{Atkins:2010yg} ignoring the background effect. However, again background 
effect can solve this issue as shown in \cite{germani2}.
In an another attempt, a non-trivial galileon \cite{Nicolis:2008in} type 
modification of higgs field has been studied, specifically by adding higher derivative operator such 
as $F(h) \partial_{\mu} h \partial^{\mu} h\Box h$ \cite{Kamada:2010qe}, in addition to the usual terms. Similar
background dependent cut off analysis has been done in \cite{Kamada}. All the aforementioned models 
are significantly constrained mainly because of nonrenormalizable interactions, that have been
introduced to get the slow roll condition. This provides tight constraints on the model parameters.
In our model we have introduced only perturbative interaction terms at the level of TeV energy scale. 
Before we go further on the theoretical description, let us provide our model predictions to the leading order
in $N$, for all the cosmologically as well as theoretically relevant quantities. As we mentioned before, considering
the efloding number to be $N = (50,60)$, we found
\bea
&&\frac{\alpha}{M_p} \simeq \left\{\begin{array}{cc} (0.0004 , 0.0003) \\ \nno
  (0.0015, 0.0014)
\end{array}\right. ~;~n_s \simeq \left\{\begin{array}{cc} (0.967, 0.972) \\ \nno
  (0.970, 0.975)
\end{array}\right. \\
&& \frac{\Delta h}{M_p} \simeq \left\{\begin{array}{cc} (0.017, 0.016) \\ \nno
  (0.21, 0.21)
\end{array}\right. ~;~r \simeq \left\{\begin{array}{cc} (2 \times 10^{-8}, 1 \times 10^{-8}) \\ 
  (9 \times 10^{-6}, 6 \times 10^{-6})
\end{array}\right. \\
&&\frac{dn_s}{d \ln k} \simeq \left\{\begin{array}{cc}(-0.0003, -0.0002) \\ 
 (-0.0003, -0.0002)
\end{array}\right. 
\eea
For the above predictions of our model, we assume a sample value for the Higgs quartic coupling $\lambda =0.2$, which
is same as its TeV scale value. If we consider the PLANCK central value of the scalar spectral index, $n_s =0.968$, with the 
same $\lambda$ value, our model predicts the number of efolding $N \simeq (52,47)$ for first and the second 
form of the potential respectively. All the other aforementioned quantities remain almost the same in terms of order of magnitude.
However, it is important to point out that because of renormalization group (RG) running, the value of Higgs quartic coupling decreases from its TeV scale value as one increases 
the energy. Keeping this in mind in Fig(\ref{lambda}), for the first form of the potential, we consider two different values
of $\alpha =0.001,0.0004$ in unit of $M_p$, and show how $(\lambda, N)$ value changes with $n_s$.
According to the present standard model calculation, it is shown that Higgs quartic coupling may hit the zero value and turn into negative
at the scale within, $10^{9}-10^{14}$ GeV, depending upon different values of Higgs mass, $125~ \mbox{GeV}~ m_h < 126~ \mbox{GeV}$, 
top mass, $m_t=(173.2 \pm 0.9)~\mbox{GeV}$ \cite{tevatron}, and
the electromagnetic coupling $\alpha_s =0.1184 \pm 0.0007~\mbox{GeV}$ \cite{bethek}. 
However, we will not discuss about this stability bound
in the current context. Of course we would like to point out that in our proposal, the original Higgs potential contains 
an infinite series of $\alpha \simeq (10^{-4}, 10^{-3}) ~M_p$ suppressed operator for the first and second forms
of the Higgs potential respectively, at TeV scale. Therefore, modification to all the aforementioned bounds need to be calculated, 
and those will be extremely important on the stability bound. Form the cosmological point of view, we see our 
model perfectly fits with the cosmological observation, see Fig.\ref{lambda}, within $ N \simeq 45- 65$ efolding number. 
Prediction of tensor mode is almost negligible within a huge range of $\alpha$, 
making it very difficult to be observed in the foreseeable future. We also point out 
that our minimal model predicts sub-Planckian field excursion, ${\cal O}(10^{-2} M_p)$, within the 
required value of $\lambda$. 
It is worth mentioning that the current PLANCK bound on running is given as $dn_s^k =-0.003\pm 0.007$ combined with the Planck lensing likelihood. Hence our predicted value is outside the PLANCK sensitivity.

{\it Inflationary energy scale and Reheating Temperature:}
As we have seen, in order to have an excellent agreement with the current observation 
by PLANCK on the inflationary observables, our minimal Higgs inflation requires a scale 
$\alpha \simeq 10^{14} \mbox{GeV}$ for the first model, and 
$\alpha \simeq 10^{15} \mbox{GeV}$ for the second form the potential, such that TeV scale quartic Higgs coupling
remains of the order of unity. Related to this prediction of $\alpha$, another 
important quantity of interest is inflationary energy scale which we defined as 
$\mu \simeq V_h^{1/4} \simeq (\lambda/4)^{1/4} \alpha \simeq (10^{14}, 10^{15}) \mbox{GeV}$ for 
both the models. It is important to notice that, our predicted inflationary energy scale is almost 
at the higher edge of the Higgs vacuum instability bound mentioned before.
This also means that Higgs mass $m_h \simeq 126$ GeV could be favourable. 
Another interesting fact is that the inflationary Hubble scale turned out to
be significantly lower than the energy scale $\alpha$
\bea
H \simeq \lambda^{\frac 12 } \left(\frac {\alpha}{M_P}\right)^2 M_p \simeq(10^{11}, 10^{12})\mbox{GeV}.
\eea
This is much lower than the naive cut off scale $\alpha$ mentioned before.
However, 
as we mentioned, unitarity has to be re-analysed in order to make any such concrete conclusion. 
    
Even though the inflationary energy scale sets the higher limit of the total energy budget of the universe,
important quantity is the amount of energy transferred from the inflaton to matter field, which
is very important for the subsequent evolution of the universe. An important part of any inflationary cosmology is 
the the mechanism of energy transformation which we call reheating phase of the universe. This is the phase when most of the inflaton
energy density is transferred to the usual matter fields which we observe today. 
This is phase which is not very clearly understood. We will not be doing detail computation at this stage.
However, an approximate estimate can be done on the upper bound of the most important quantity called 
reheating temperature, $T_{re}$, based on the assumption 
that the reheating, and the subsequent thermalization happens after the inflation because
of the interactions of the Higgs boson with the standard model particles.
The maximum possible transferred matter energy density $(\rho_{m})$ after the reheating phase 
can be identified with the inflaton energy density $(\rho_{h_{end}})$, when $\epsilon(h_{end}) \simeq 1$, is satisfied.
This is the condition for the end of inflation, we have considered before. 
Hence, one can arrive at following equality 
\be 
 \rho_m \simeq \rho_{h_{end}} \simeq 2 V_{h_{end}}.
\ee 
Energy transfer and thermalization of the matter field generally are not the instantaneous process.
Therefore, equilibrium temperature which mentioned before as a reheating temperature, $T_{re}$, 
of the produced relativistic matter field will be in general smaller than the maximum value attainable.   
In thermal equilibrium the energy density of a relativistic matter filed can be expressed in terms of equilibrium temperature
as, 
\bea
\rho_m=\frac{g_*\pi^2T_{re}^4}{30}~,\implies T_{re} \lesssim  \left\{\begin{array}{cc} 10^{13} \mbox{GeV}  \\ \nno
  10^{14} \mbox{GeV} ,
\end{array}\right.
\eea
where $g_*\simeq 106.75$ is the numbers of relativistic degree of freedom during reheating.
Therefore, maximum reheating temperature will well below the Planck scale. This is consistent with the constraint from Big Bang
Nucleosythesis.

{\it Background dependent unitarity}: As mentioned before standard power counting analysis tells us that the
tree level unitarity scale $\Lambda_{tree}$ should be of the order of same as inflationary energy scale $\alpha$.
It is already well established that in standard model unitarity is dependent upon the 
Higgs vacuum expectation value. Therefore, this should also be true in the present context. From general
power counting argument, the background dependent cut off scale $\Lambda(h)$ 
can be read off from the coefficient of the operator of dimension higher than 
four. In our minimal Higgs inflation scenario, all the higher dimensional operators come from the expansion 
of $V_h$ in the inflationary background $h_0$ by expanding $ h =h_0 + \delta h$. After expansion, the five dimensional 
operator turned out to be
\bea
\delta V_h = \left\{\begin{array}{cc}
 \frac {8 \lambda {\tilde h}_0^3 (-7 + 57 {\tilde h}_0^4 - 57 {\tilde h}_0^8 + 7 {\tilde h}_0^{12})} {\alpha (1 + 
   {\tilde h}_0^4)^6} \delta h^5 \\ 
 \frac{12 \lambda {\tilde h}_0 (-1 + 7 {\tilde h}_0^2 - 7 {\tilde h}_0^4 + {\tilde h}_0^6)}{
 \alpha (1 + {\tilde h}_0^2)^7} \delta h^5 \end{array}\right.
\eea
Therefore, at large field limit ${\tilde h} \gg 1$ (inflationary regime), one finds 
\bea
\delta V_h = \frac 1 {\Lambda(h)} \delta h^5 \implies \Lambda(h) \simeq \left\{\begin{array}{cc}
\frac{\alpha}{56 \lambda} {\tilde h}^9 \\ \nno
\frac{\alpha}{12 \lambda} {\tilde h}^7 \end{array}\right.
\eea
This is much larger than the tree level unitarity limit $\alpha$. Therefore, 
our inflationary predictions could be robust against quantum corrections. 
We defer the detailed unitarity analysis for future study. 
However, interesting point to mention, some odd behaviors are turning up for $ 0.4 < \tilde h < 2.5$ because of 
the presence of zeros of $\Lambda(h)$ for some specific values of $\tilde{h}$. It could be interesting to understand this peculiar
behavior.  

{\it Discussions and Conclusion.} In this Letter we have proposed a new 
Higgs inflation scenario, where, the field is minimally coupled with gravity.
In the usual Higgs inflation, a non-minimal coupling of the Higgs field with
the gravity is introduce. The non-minimal coupling keeps the Higgs potential 
sufficiently flat by increasing the effective Higgs field dependent 
gravitational coupling. In our minimal Higgs inflation scenario,
we modified the potential by introducing a new scale $\alpha$ in such a way that it creates a large flat plateau for the potential 
at the inflationary energy scale $\mu \sim \lambda^{1/4} \alpha$. Therefore, our modified Higgs potential naturally contributes an 
infinite series of higher dimensional operators at the TeV scale, however, we expect the contribution of those
operators on the usual standard model observables to be significantly suppressed by the 
aforementioned inflationary energy scale $\mu$. It is therefore, important to check how, the coupling constants
of those infinite series of higher dimensional operators flow towards the TeV scale under RG flow. 
By now we understood the fact that all the non-minimal Higgs inflationary models are plagued by unitarity issue, or is tightly constrained.
Therefore, alternative scenarios are welcome in this regard. As we have explained through out
this paper, our minimal Higgs model fits extremely well with all the observations made by PLANCK related
to the inflationary dynamics. More specifically, for both form of the Higgs potential, prediction of tensor to scalar ratio turned out to be
very small ($r < 10^{-5}$), which is very difficult to observe in near future. It is also worth pointing out that because of 
the minimal single field inflationary scenario, the non-gaussianity will also be very small \cite{maldacena}. For both the form of the potentials
the value of scalar spectral index turned out to be $n_s = (0.967, 0.972)$ considering $N=50$. 
We also the computed the background field dependent unitarity scale $\Lambda(h)$, which turned out to be
much larger than the tree level unitarity scale $\alpha$. Currently we are working on more detail analysis 
of our models from the cosmology and the particle physics point of view.

{\bf Acknowledgement}\\
We are very thankful to our HEP-group members to have vibrant academic discussions.

\end{document}